# High-strength cellulose–polyacrylamide hydrogels: mechanical behavior and structure depending on the type of cellulose

A.L. Buyanov*, I.V. Gofman, N.N. Saprykina

Institute of Macromolecular Compounds, Russian Academy of Sciences, Bolshoy pr., 31, St-Petersburg, Russia

* Corresponding author. E-mail: buyanov@hq.macro.ru

Two types of high-strength composite hydrogels possessing the structure of interpenetrating polymer networks were synthesized via free-radical polymerization of acrylamide carried out straight within the matrix of plant or bacterial cellulose swollen in the reactive solution. The mechanical behavior of synthesized hydrogels subjected to the action of compressive deformations with different amplitude values was studied. The analysis of the stress-strain curves of compression tests of the hydrogels of both types obtained in different test conditions demonstrates the substantial difference in their mechanical behavior. Both the plant cellulose-based and bacterial cellulose-based hydrogels withstand successfully the compression with the amplitude up to 80 %, but the bacterial cellulose-based compositions demonstrate in the multiple compression tests the substantial decrease of the mechanical stiffness caused by the action of compressions. This effect takes place straightly during the multiple cyclic compression tests or in the course of relaxation of previously compressed samples in water in unloaded state during 2-3 days after the first compressive tests. Being subjected to the action of the extremely sever compression (up to 80 %) the bacterial cellulose-based hydrogel varies greatly its mechanical behavior: during the subsequent compressions the stiffness of the material keeps extremely small (as compared to that registered in the first compression) up to the deformations as high as 50-60 % with the dramatic increase of the stiffness during the further loading. This unusual effect can be explained by the deep reorganization of the intermolecular structure of the material with the stress-induced reorientation of cellulose micro-fibrils, fragmentation and disintegration of the interpenetrating networks structure with the breaking of the covalent bonds in the matrix polymer. Submicron- and micron-scale specific features of structures of composite hydrogels of both types were studied by cryo-scanning electron microscopy to explain the peculiarities of observed mechanical effects.

### Introduction

Hydrogels are the polymer networks that are swollen in water and exhibit elastic properties. Some of them resemble living tissues in their physical properties and have good biocompatibility, which allows them to be considered as potential substitutes of soft human tissues, in particular, of cartilage tissues and vessels (Ku, 2008, Batista et al., 2012, Sciaretta, 2013, Gonzales, Alvarez, 2014). Currently, however, the problem of improvement of the mechanical properties of these hydrogels up to the level characteristic of natural articular cartilage has not been solved completely (Batista et al., 2012, Gonzales, Alvarez, 2014), and this situation hampers the wide use of hydrogels in the appropriate domain of medicine.

In recent years, cellulose, in particular the specific type of this polymer, namely bacterial cellulose (BC), has been used for the synthesis of various types of advanced composite materials including hydrogel compositions for biomedical applications (Figueiredo et al., 2013, 2015, Lin, Dufresne, 2014, Nandgaonkar et al., 2016, Ullah et al., 2016). Cellulose is a natural polymer characterized by the high mechanical stiffness and excellent biocompatibility (Sannino et al, 2009). The composite materials with improved properties can be obtained while using cellulose as a component of compositions with other polymers.

As a rule, cellulose in different forms is introduced into synthetic hydrogels to improve their mechanical properties. For example, just some little amounts of micro-granulated cellulose being introduced into the hydrogels of polyacrylamide (PAAm) at the stage of their synthesis (10 - 150 mg per 1 g of acrylamide) insure the increase of the Young's modulus of the material up to 1.5 times (Sahiner, Demirci, 2017).

But the most substantial improvement of the mechanical properties of hydrogels can be obtained in the composite materials with the structure of interpenetrating networks (IPN). We have used the method of IPN-hydrogels synthesis to develop the composite materials consisting of cellulose-PAAm and cellulose-polyacrylic acid (PAA) compounds (Buyanov et al., 1998, 2001, 2010, 2013). Due to the high rigidity of cellulose chains, these hydrogels possess the improved values of mechanical strength and stiffness, and at the same time retain all valuable properties inherent to PAAm and PAA. As a result these hydrogels exhibit high stiffness, strength, and flexibility under different types of mechanical loads, including long-term multiple cyclic compressions (Buyanov et. al, 2010, 2013, Gofman, Buyanov, 2017). In the synthesis of our hydrogels, we used cellulose of two types as the reinforcing component, namely BC or regenerated plant cellulose (PC). It was shown in the *in-vivo* experiments with the laboratory animals that both types of hydrogels are characterized by a perfect biocompatibility and can work during long periods of time in the animals' joints as the artificial substituent of cartilage (Buyanov et al., 2016). During these tests, the high level of the integration of hydrogel implants with surrounding living tissues was evidenced. In particular, the formation of a "shell" of calcium phosphate spherulites in the hydrogel volume near the implant-bone interface has been registered and this effect insures the mechanically strong connection of the implants with the bone tissues, which is very important for practical use of these materials as artificial cartilage tissues (Buyanov et al., 2016, Gofman et al., 2018).

As it was shown in our investigations the mechanical behavior of hydrogels containing two types of cellulose, BC or PC, can differs significantly. They are characterized by different stiffness values and by different ability to withstand the high deformations, in particular, the compressive deformations in the conditions of both one-shot and cyclic tests (Buyanov et al., 2010, 2013, 2016, Gofman, Buyanov, 2017).

One more important feature of the mechanical behavior inherent to the BC-PAAm hydrogels should be stressed, that was evidenced namely in our previous work, i.e., an anisotropy of mechanical properties, detected during compression (Buyanov et al., 2010).

The flat sheets of swollen BC used in our work as the matrices to form the hydrogel samples were cultivated by the life activity of Acetobacter bacteria which start to produce the 3D physical network of the polymer at the surface of nutrient solution. All mechanical characteristics of BC-PAAm hydrogel samples were found to be considerably higher when measured along the direction perpendicular to the top and bottom surfaces of the original BC sheets, than those registered in the direction parallel to these surfaces (Buyanov et al., 2010). It was suggested that the anisotropy of mechanical properties of BC-PAAm hydrogels is associated with the structural features of BC. According to Thompson et al. (Thompson et al., 1988) there are some ''tunnels'' in the BC volume oriented mainly in the vertical direction, in the direction of thickness of the BC layer: bacteria form these tunnels during biosynthesis. The ''walls'' of such tunnels are presumably formed by the rigid microfibrillar BC ribbons, and they are able to insure the reinforcement of hydrogels under compression in the vertical direction. In our case, tunnel lacunas in the BC structure are filled with relatively soft PAAm chains that results in the decrease of compressive stiffness in the direction along the BC surface (Buyanov et al., 2010).

Since PAAm is a common component of both BC- and PC-based hydrogels (the concentration of this component in the dry cellulose-PAAm composition is as high as 80-95 %), the aforementioned differences in the mechanical properties of these materials, obviously, originate only from the different characteristics of the mesostructure of two different types of cellulose used in the synthesis, BC or PC. The results obtained at the initial stage of the investigation of the structure of anisotropic BC-PAAm hydrogels were presented in our previous work (Velichko et al., 2017). The simultaneous use of the methods of spin-echo small-angle neutrons scattering (SESANS) and SEM-techniques equipped with cryo-camera (Cryo-SEM) reveals the existence of sizable tunnel-like structural inhomogeneities of the characteristic transversal dimensions more than 10 μm in the volume of swollen BC-PAAm hydrogels. The work described below aimed at a careful comparison of the characteristics of the mechanical behavior of the hydrogels of both types, namely those based on BC and PC, and at the clarification of the correlations between the differences in their mechanical properties and specific features of their structures. It was thus of interest to characterize the mechanical behavior of the hydrogels under study in the conditions of both one-shot and multiple cyclic compression acts up to the extremely high deformation value, up to 70 - 80 %. Of especial interest was to study the mechanical behavior of these composite hydrogels in the conditions of cyclic compression tests, because this mode of loading will take place in practical use of the materials under investigation. Such data can hardly be found in the literature.

Since the structure of the composite hydrogels based on PC was not previously studied, it was necessary to obtain the information concerning this subject to correlate the mechanical properties of the hydrogels with characteristic features of their structure. To get this goal the Cryo-SEM method was used like in (Velichko et al., 2017), which allows studying water-swollen soft materials maintaining the sample as close as possible to its natural state because the severe collapse and flattening due to dehydration does not exist in properly frozen specimens (Apkarian, Wright, 2005).

**Materials and methods**

Acrylamide purchased from Sigma-Aldrich Corp. was recrystallized twice from benzene. All other chemicals of analytical grade were used as received. To prepare the resulting hydrogel samples the swollen gel-like cellulose matrices were used in the form of flat sheets of the thickness of up to 15 mm.

The composite hydrogels were synthesized using the technique that we have developed previously, namely by radical copolymerization of acrylamide with the low molecular weight crosslinking agent, N,N'-methylene-bis-acrylamide (MBA) conducted inside the cellulose matrices swollen in the reaction solution (Buyanov et al., 1998, 2010, 2013, 2016).

The initial concentrations of acrylamide, MBA and cobalt (III) acetate, used as the initiator, in the water solution were 7.4, $1.4 \times 10^{-3}$ и $0.5 \times 10^{-3}$ mol/l, respectively. The extent of the monomer conversion close to 100 % was obtained after 2 h of the reaction at 25 $^O$C. Then the composite hydrogels were placed in distilled water for several days to remove the residual amounts of low molecular weight components and to let the gels swell.

For our investigations the PC-PAAm and BC-PAAm hydrogel samples with approximately same equilibrium water contents and the same levels of initial stiffness were chosen, which can withstand the compression up to 80 % without the visible symptoms of destruction and fragmentation.

The water content in the samples in the equilibrium state was determined by drying of thin hydrogel plates at 160 $^O$C up to the constant weight value. The equilibrium water contents in the BC-based and PC-based samples were found to be 70 and 68 wt%, respectively. Quantitative chemical composition of the samples was précised both by the elemental analysis and by the gravimetric method basing on the amounts of the components introduced in the reaction. The contents of cellulose in the BC-PAAm and PC-PAAm compositions were found to be 3.5 and 5.0 wt%, respectively.

The hydrogel samples were prepared in the form of the flat plates. To test the mechanical properties of the materials under study the cylindrical specimens with the diameter of 7-10 mm and thickness 4-5 mm were cut from these plates in two directions: perpendicular to the surface of the hydrogel plate and parallel to the surface. The mechanical characteristics measured in these directions were designated in the text below by the symbols ⊥ and □, respectively.

The mechanical tests of the materials under study have been carried out at the AG-100kNX Plus mechanical test system (Shimadzu Corp., Japan) predominantly in two test modes, namely single-shot compression with the constant speed of 10 mm/min up to the preselected deformation value, as a rule up to the compression as high as 80 % (the initial tests) and multiple cyclic compression tests with the same speed up to the preselected deformation amplitude value (these tests aimed to determine the stability of the mechanical behavior of the studied hydrogels).

The single compression tests were carried out in air medium taking into account the short duration of these tests: the model experiments have shown no marked loss of water by the sample during these tests. The long-term cyclic tests were performed in water medium.

During these tests we first of all have registered the stress-strain curves, the dependences of the compressive stress $\sigma$ upon the deformation $\varepsilon$ of the samples. To characterize the stiffness of the hydrogels under tests the mean slope values of the compression curves $\Delta\sigma/\Delta\varepsilon$ were determined in two ranges of the compressive deformation: from 10 to 15 % and from 25 to 30 % ($E_{10-15\%}$ and $E_{25-30\%}$, respectively). Three stress values were additionally determined in the single compression tests: those corresponding to the compression values of 30, 50, and 80% ($\sigma_{30\%}$, $\sigma_{50\%}$ и $\sigma_{80\%}$). In the cyclic tests, the stress values corresponding to the amplitude compression values were determined.

Cryo-SEM was performed on a Supra 55VP scanning electron microscope (Zeiss, Germany) equipped with a PP 2000T Cryo-SEM system (Quorum Technologies, the United Kingdom). The hydrogel samples swollen to the equilibrium state were rapidly frozen in liquid nitrogen and then transferred to preparation chamber cold stage (-140°C) and fractured. The sample chips were oriented so that their surface focused mainly perpendicular to the surface of growth of matrix BC. For improving contrast, the sample surface was coated with platinum by cathode sputtering. For ice sublimation, etching of the surface and revealing structure, the temperature of the cold stage was raised to −90°C. After coating with platinum, the sample transferred to the cold SEM stage (-140°C). Independent cooling of the cold stage and the cold trap (-190$^{O}$C) prevented ice buildup on the sample. The surface morphology was studied using the secondary electron mode (SE2).

### Results and Discussion

The stress-strain curves of the composite hydrogel BC-PAAm tested in the single compression mode in two directions (perpendicular and parallel to the BC matrix plate surface) and of the hydrogel PC-PAAm tested in the same mode are demonstrated in Fig. 1. The mechanical characteristics of the materials are listed in Table 1.

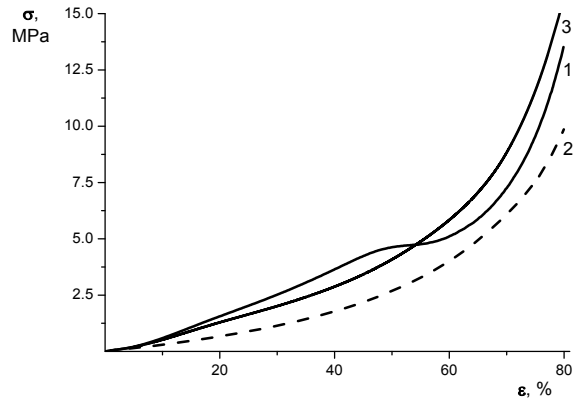

Fig. 1. Stress-strain curves obtained during the compression tests of the (1, 2) BC-PAAm hydrogel samples and (3) PC-PAAm hydrogel sample; the curves of the compression ob BC-PAAm sample in the directions (1) perpendicular and (2) parallel to the top and bottom surfaces of the original BC sheet are presented.

Table 1

Mechanical characteristics of the hydrogel samples determined in two consecutive single-shot compression tests up to the deformation of 80 %. The properties registered during the first and the second compression acts are designated by the figures 1 and 2.

| Gel type | Test conditions | $E|_{10-15\%}$, MPa | $E|_{25-30\%}$, MPa | $\sigma|_{30\%}$, MPa | $\sigma|_{50\%}$, MPa | $\sigma|_{80\%}$, MPa |
|---|---|---|---|---|---|---|
| BC-PAAm | 1-⊥ | 9.43 | 9.15 | 2.52 | 4.64 | 14.6 |
| BC-PAAm | 1-∥ | 3.49 | 4.84 | 1.15 | 2.69 | 9.83 |
| BC-PAAm | 2-⊥ | 0.56 | 1.17 | 0.26 | 1.05 | 76 |
| BC-PAAm | 2-∥ | 0.47 | 0.69 | 0.18 | 0.51 | 16.3 |
| PC-PAAm | 1 (⊥ and ∥) | 7.76-7.69 | 7.27-7.34 | 1.98-1.92 | 4.11-4.17 | 15.7-15.3 |
| PC-PAAm | 2 | 1.73 | 2.41 | 0.56 | 1.27 | 15.4 |

The first thing that should be stressed while analyzing the obtained results is a pronounced anisotropy of the mechanical properties of the BC-based hydrogel. The values of all mechanical characteristics of the material, determined in the compression along the axis perpendicular to the surface of the hydrogel plate (Fig. 1, curve 1), exceed the values of the same characteristics, measured in the direction parallel to this surface (Fig. 1, curve 2).

Another situation was evidenced while testing the PC-based hydrogel sample: no appearance of the anisotropic mechanical behavior was detected. The same values of the properties of this material were detected while compressing the samples in all directions (Table 1).

While comparing the properties of the hydrogel of different types measured in this single compression test no substantial difference was detected of the characteristics of PC-based hydrogel and of BC-based one measured in the ⊥ direction.

The hydrogel samples of both types successfully withstand the high compressional deformation (Fig. 1). Some plateau can be seen at the curve 1 (BC-PAAm gel compressed in the ⊥ direction) in the deformation range about 50 %. According to the literary data (Stammen et al, 2001), the appearance of this feature of the curve points to the onset of transformation of the hydrogel's structure. In general, this transformation can cause the destruction of the material (Stammen et al, 2001). But for the hydrogel under study no visually seen indications of the sample's destruction were detected after the compression of both PC-based sample and of BC-based specimen (both in ⊥ and ∥ directions) up to the maximal compression value as high as 80 %.

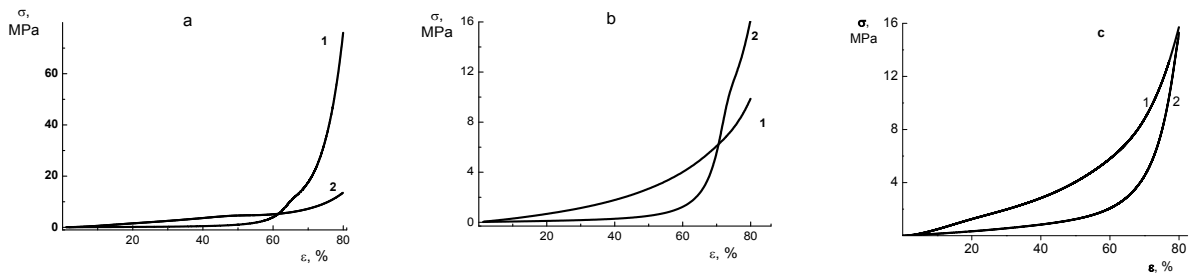

Fig. 2. Stress-strain curves obtained during two successive compression tests of (a, b) the same BC-PAAm and (c) PC-PAAm hydrogel samples: compression of BC-PAAm samples in the directions perpendicular (a) and parallel (b) to the surface of matrix BC layer; 1 - first compression, 2 - second compression after 30 min. of relaxation in water.

Nevertheless after so high compression the hydrogel samples become substantially softer than in the initial state. Fig. 2 presents the stress-strain curves of the repeated compression of the samples under test after a 30 min. period of relaxation in water in unloaded state. These curves are compared with those obtained during the first compression tests, and the appropriate mechanical characteristics of the samples measured in both first and second tests are presented in Table 1.

The samples, previously subjected to an extensive compression up to 80 %, demonstrate during the second compression the stress values (in the deformation range up to 40-50 %), which are up to one order of magnitude less than those in the first compression.

A different character of the mechanical behavior is inherent to the BC- and PC-based gels during the further stage of the second compression (from 50-60 to 80 %). The fast increase in the compression stresses was registered while studying the BC-PAAm gel. The stress registered in this deformation range during the second compression exceeds this realized in the first compression (the most dramatic stress increase up to 76 MPa takes place in case of the deformation of this gel sample in the $\perp$ direction (Table 1)). For the PC-PAAm sample the stress value in the second test increases rapidly too in this deformation range, but up to the maximal compression realized in these tests (80 %) the stress values remain less than those registered in the first compression (Fig. 2c, Table I). No effects similar to those registered in the same conditions while testing BC-PAAm samples (Fig. 2a, 2b, Table I) were detected in this case.

While treating these results one can suppose that in both BC-PAAm and PC-PAAm hydrogels the deep re-orientation of the micro-fibrillar structure of cellulose carcass of the hydrogel composition takes place in the conditions of the extensive compression (up to 80 %). Probably, the partial destruction of the covalent bonds of the cellulose network structure can take place, but this event does not provoke the formation of cracks and any other visible defects of the gel structure. It is probable also that the rearrangement of the system of inter- and intra-molecular hydrogen bonds within IPN structure of hydrogels and, presumably, simultaneous rearrangement of the fluctuation network of entanglements between the polymer components of IPN occur (Buyanov et al., 2013).

But presumably in PC-PAAm hydrogels the changes of the cellulose network structure are not as extensive as those in BC-PAAm gels. As a result the tests of PC-PAAm samples does not show the huge increase in the compressive stress values during the second compression in the high deformation region that was registered while testing the BC-PAAm samples (Table 1). Really, the intermolecular structure of the fibrils packing in PC is not so developed and complicated as the structure of BC that may cause the difference in the above described peculiarities of the behavior of these two types of hydrogel compositions.

While analyzing the stress-strain curves of the compression tests of PC-PAAm samples, no plateau can be seen similar to that inherent to BC-PAAm gel in the deformation range about 50 %: the stress value increases monotonically up to the maximal compression values realized in our tests. Apparently this difference reflects the difference in the characteristics of intermolecular organization of BC and PC matrices of the hydrogels' composition. The structure of BC-PAAm composition is more developed than that of PC-PAAm that provokes the realization of multi-stage process of the destruction of BC-PAAm hydrogel under the action of the compressive load.

It is important to note that the above mentioned anisotropy of the mechanical properties inherent to the BC-PAAm hydrogels was registered during the repeated compression too: all characteristics determined in the compression, directed perpendicularly to the BC surface, substantially exceed the appropriate values measured while compressing in the direction parallel to the surface (Table 1).

In our previous works the described anisotropy was hypothesized to be caused by the specific features of the BC structure (Buyanov et al, 2010). Indeed, according to the data presented in (Thompson et al., 1988) this structure is characterized by the presence of a system of "tunnels" that are oriented in the direction transversal to the surface of the BC layer. These tunnels are formed by the bacteria during the BC cultivation. The "walls" of these tunnels may be reinforced by the agglomerates of rigid-chain micro-fibrillar tapes of BC that would virtually strengthen the material in the appropriate direction (perpendicular to the BC sheet surface).

On the other hand in composite hydrogels studied these tunnels must be filled with comparatively soft chains of PAAm. This circumstance will cause the decreased stiffness of the material while being compressed in the direction parallel to the BC surface (Buyanov et al, 2010). Under the hard compression of the hydrogel samples such oriented structure may be subjected to some transformations and to a partial destruction.

The characteristic that is most important for successful practical use of the hydrogels under study as the joint cartilage implants is their ability to withstand the multiple actions of the cyclic compressions as high as 50 %: it is a maximal compression value registered in the normal mode of human cartilage function (Ker, 1999).

But the study of the behavior of these materials under the action of yet higher compressive loads is of interest too. This information is evidently necessary to determine the limit levels of the loads and deformations that these materials can withstand without the substantial depression of their principal functional properties.

Fig 3 presents the stress-strain curves, obtained in the conditions of multiple cyclic compression (1000 cycles) of both BC-PAAm and PC-PAAm samples with the deformation amplitude as high as 30 % (according to the data obtained while testing the joint cartilage (Stammen, 2001), this is a typical value of compression amplitude of cartilage tissue of human knees and hip-joints tor the persons with the typically middle level of motional activity).

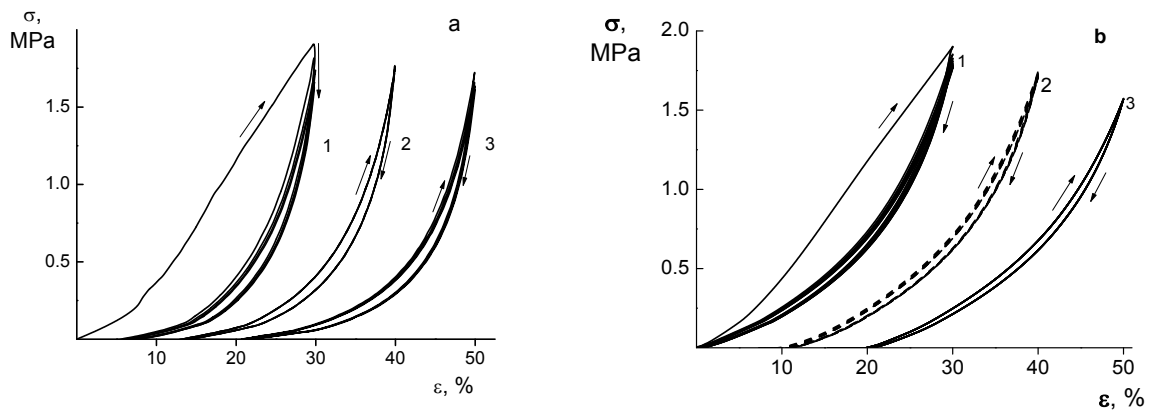

Fig. 3. Stress-strain curves of multiple cyclic compressions of (a) BC-PAAm and (b) PC-PAAm hydrogel sample (1000 cycles, compression amplitude - 30 %). 1 - cycles 1-10, 2 - cycles 500-510, and 3 - cycles 990-1000. While testing the BC-PAAm sample the compressive load was directed perpendicularly to the surface of matrix BC. For convenient data presentation each group of cyclic curves in this and following figures was shifted at 10 % along the deformation axis against the previous one.

The cyclic stress-strain curves obtained are characterized by a well-defined hysteresis that is inherent for the cyclic deformation of both the cartilage and of the hydrogels studied, which are very close to the cartilage by their stiffness (Buyanov et al., 2013). A most broad hysteresis is characteristic for the first compression cycle while already in the second and following cycles the substantial depression of the hysteresis area can be seen (Fig. 3): the ratio of the areas of the first and second cycles is 5.66 for the BC-PAAm hydrogel sample and 4.63 for the PC-PAAm one. The area of cycle remains practically constant from the second cycle up to the end of the experiment (up to 1000-th cycle).This feature of the cyclic compression behavior is characteristic for the hydrogels of both types. The most important feature of the material's behaviour in these conditions is that up to 1000-th compression cycle no appreciable decrease of the maximal stress value, corresponding to the amplitude compression, was registered.

Analyzing the results obtained in this experiment one can conclude that under the action of moderate compressions (about 30 %) no substantial difference in the behavior of these two materials can be registered (compare Figs. 3a and 3b).

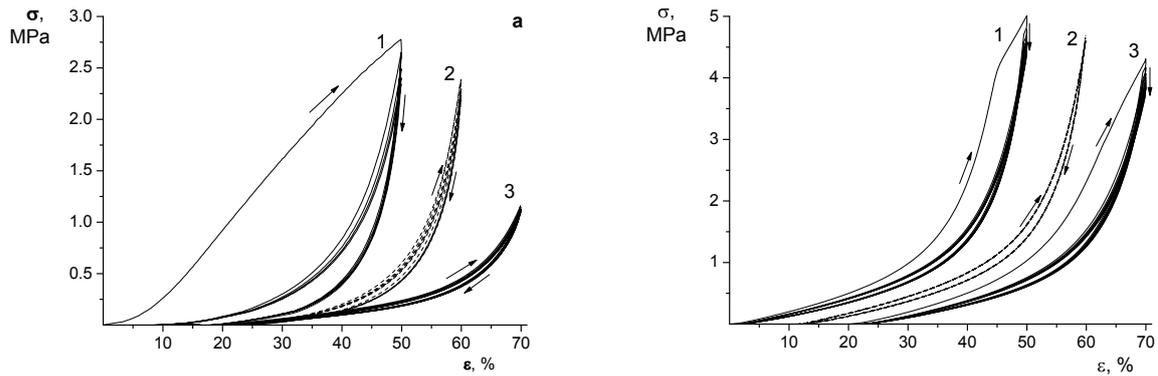

Fig. 4. Stress-strain curves of multiple cyclic compression of (a) BC-PAAm, and (b) PC-PAAm hydrogel samples, compression amplitude - 50 % (100 + 100 cycles with the relaxation of the samples in water during 2 days after the first series of compressions). 1 - first series, cycles 1-10, 2 - first series, cycles 90-100; 3 - second series (after relaxation), cycles 1-10. While tested the BC-PAAm sample the compressive load was directed perpendicularly to the surface of matrix BC. For convenient data presentation, groups 2 and 3 of cyclic curves were shifted at 10 % along the deformation axis against the cycles of previous group.

Fig. 4 presents the stress-strain curves obtained while cycling twice the samples of the same hydrogels, namely BC-PAAm and PC-PAAm up to the compression amplitude of 50 %. An unusual effect was registered in these experiments while testing the BC-PAAm sample (Fig. 4a): the sample that was subjected to the action of multiple (100 cycles) cyclic compressions with the amplitude of 50 % (fig. 4a-1, 2) and then kept in water in the unloaded state for 2 days, when subjected to the repeated cyclic compression in the same conditions demonstrated the substantially depressed level of the mechanical stiffness (fig. 4a-3). The maximal compressive strength corresponding to the amplitude value of deformation of this sample (50 %) registered during the second series is about 1 MPa while in the first series of compression cycles it was as high as 2.4-2.8 MPa. We should note that the control experiment carried out in the same conditions but with the compression amplitude of 30 % did not show the same effect: the amplitude strength registered in the second series of compression cycles did not differ significantly from the value obtained in the first series.

This effect was not obtained while the PC-based gel sample was tested in the same conditions (Fig. 4b): the stress values registered in the second series of cycles (after relaxation during 2 days, Fig. 4b-3) does not differ significantly from those registered in the first series (Fig. 4b-1, 2). This difference in the mechanical behavior of two types of hydrogels reflects the substantial difference in their structures.

One can conclude that, while increasing the compression amplitude of the BC-based gel sample up to 50 %, some sizable reformations of the material's structure have apparently happened as a result of the action of the first series cyclic loadings. The break of the initial system of covalent bonds of the interpenetrating network can be hypothesized.

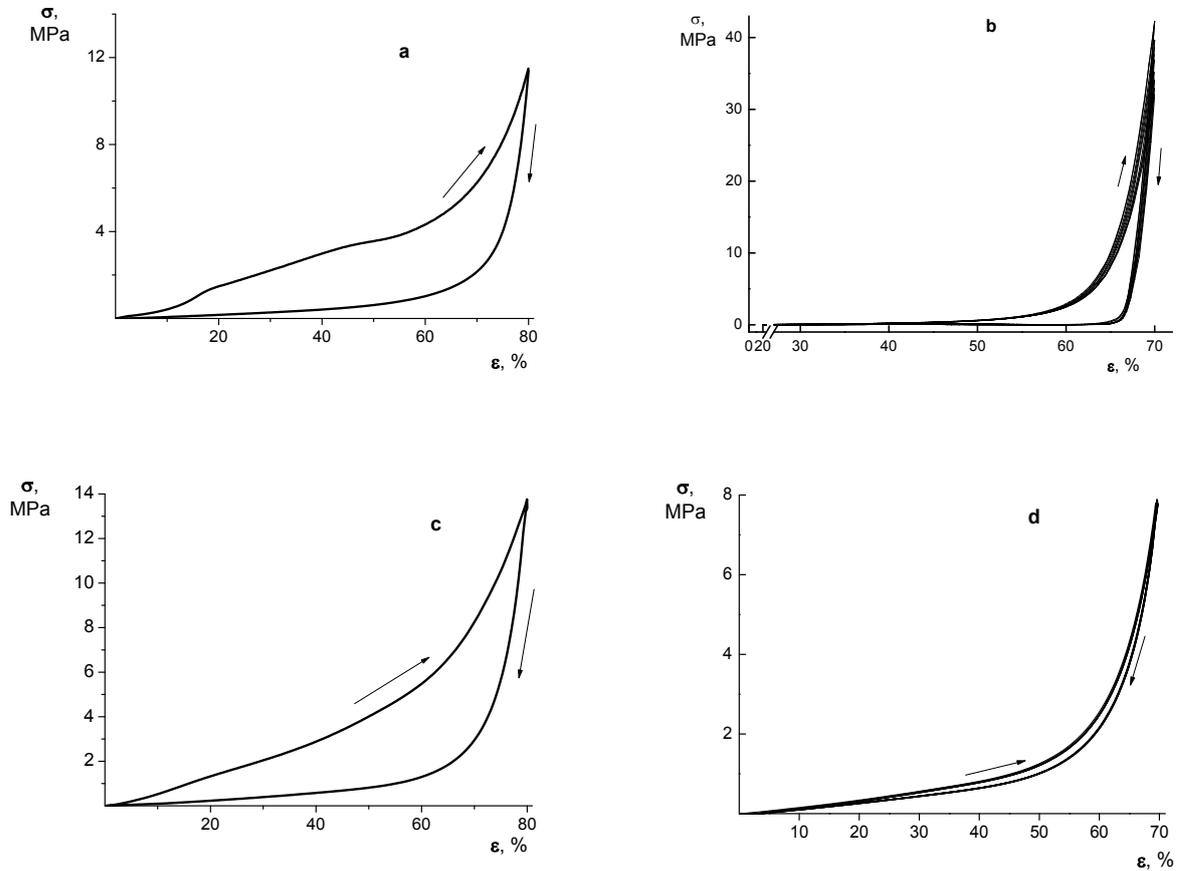

Fig. 5. Stress-strain curves of cyclic compression of (a, b) BC-PAAm and (c, d) PC-PAAm hydrogel samples up to 80 % (1 cycle – a, c), followed by multiple cyclic compressions (10 cycles – b, d) of the same samples with the amplitude of 70 %.

It is important to note that these deep variations of the mechanical strength was not detected just during the first series of cycles itself (fig. 4a-1, 2), but were evidenced only after long-time relaxation of the sample in water in unloaded state (fig. 4a-3). Probably, the action of the heavy cyclic loading provoked the rupture of the integrity of the interactions between the physical network of BC and chemical network of PAAm. This effect may result from the deep differences in the stiffness of PAAm and cellulose macro-chains (Buyanov et al., 2013). One can conclude that these processes of the reorganization of the network structure take a substantial time. Their results appeared not just during the first series of cycling, but only after a substantial period of time.

The yet more pronounced differences in the mechanical behavior were detected under the further increase of the compression amplitude of the gel samples studied. In these conditions not only the decrease in the stiffness of the BC-based material after the cyclic loading takes place without the change of the shape of the curves, but the qualitative variation of their form is registered too.

Fig. 5 presents the results of the experiment in which both (a, b) BC-PAAm and (c, d) PC-PAAm gel samples were primarily subjected to the action of cyclic compression up to an extremely high compressive deformation of 80 % (a, c) and then, after a 30 min period of relaxation in water, they were cyclically compressed (b, d) 10 times with the amplitude of 70 %. While testing the BC-based sample the load in these tests was directed perpendicularly to the surface of the hydrogel plate.

It can be clearly seen in Fig. 5b that the action of this intense preliminary compression provokes the deep changes in the mechanical behavior of the BC-based material. In the second (after a relaxation) series of compression cycles the compressive stress value in the range of the deformations up to 45-50 % remains almost constant and very low, and the hydrogel's stiffness is kept at a low level: the mean value of the compressive modulus $\Delta\sigma/\Delta\varepsilon$ in the deformation range 40-50 % is almost one order less than that measured in the initial compression cycle of the same sample (compare Figs. 5a and 5b). But in the course of the further increase of the deformation – beyond 55-60 %, the drastic increase of the stress is registered: while increasing the compression up to 70 % (Fig. 5b), the compressive stress value was found to be about three times as much as that registered in the first compression (Fig. 5a). A well-defined hysteresis is inherent to the cyclic stress-strain curves obtained in these tests, and the maximal compression stress corresponding to the amplitude deformation tends to decrease progressively from cycle to cycle.

Apparently, in these test conditions the predominantly elastic reaction to the compression is inherent to the hydrogel only while the deformation exceeds 55-60%. And the extremely high compressive stress values registered in these tests confirm the realization of previously hypothesized process of the orientation of BC micro-fibrils along the compression direction.

The results of the tests of the PC-PAAm sample (Fig. 5c,d) carried out under the same protocol that those of the BC-PAAm gel demonstrate the deep difference in the mechanical behavior of BC- and PC-based hydrogels under the action of high compression. In these conditions the BC-PAAm gel demonstrates the extremely low stiffness in the deformation range up to 45-50 % and the effect of substantial hardening of the material under the action of high compression. The stress values corresponding to the amplitude compression in the cycles presented in Fig. 5b are in the range of 30 – 40 MPa. Another character of the cyclic stress-strain curves is inherent to PC-PAAm sample. A sizable increase of the stress along with the increase of the deformation takes place already in the low deformation range, from the very beginning of the compression. The preliminary compression up to 80 % leads for some hardening of PC-PAAm sample but the extent of this hardening is more modest as compared to this in BC-PAAm gel (compare the amplitude stress values presented in Figs. 5b and 5d).

### *Cryo-SEM data*

It would be plausible to connect the differences in the mechanical behavior of the composite hydrogel samples based on two types of cellulose with the specific peculiarities of the stricture, the morphology of these materials. The structure of these hydrogels is substantially affected by the intermolecular structure of the cellulose scaffold of the material, namely of BC in one case or of PC in another one.

To clarify this question we tried to obtain the comparative data characterizing the structure of the hydrogels under study by Cryo-SEM methods. This task was especially important while characterizing the structure of PC-PAAm, which was not studied in our previous works.

As it was mentioned above (section ''Materials and methods") for BC-PAAm we obtained the cryo-SEM images of the samples fractured thru the plane mainly perpendicular to the top and bottom surfaces of the original BC sheet.

The Cryo-SEM micrographs of composite BC-PAAm and PC-PAAm gel samples are presented in Fig. 6.

Analyzing the data presented in Fig. 6 one can see the substantial inhomogeneity of the structure of BC-PAAm hydrogel. The micro-photo with modest magnification (2 000x – Fig 6 a) demonstrates two regions of extended inhomogeneity (they are designated by ellipses); the distance between them is about 20 μm. As far as it was shown elsewhere (Velichko et al., 2017), the similar inhomogeneities can be detected in BC-PAAm hydrogels by SEM as the traces of the tunnel-like formations existing in the volume of matrix BC. They are oriented predominantly along the direction perpendicular to the surfaces of the BC sheets. The SEM examination of the structure of BC has shown a very low concentration of cellulose microfibrils inside these tunnels. The "walls" of the tunnels are formed by the condensed cellulose micro-fibrils, whose density in these regions substantially exceeds the mean density of the fibrils in the BC volume (Velichko et al., 2017). In BC-PAAm gels, the tunnel regions are filled predominantly by PAAm, which fills as well the regions of the "walls" themselves situated around the BC ribbons.

This type of anisotropic mesostructure of the BC-PAAm gel's IPNs with the density inhomogeneities of the diameters more than 10 μm was observed not only by Cryo-SEM, but also by complementary method of spin-echo small-angle neutron scattering (Velichko et al., 2017). Moreover, the Cryo-SEM data has directly shown the presence of the tunnel-like structures inside the swollen BC matrix (Velichko et al., 2017).

Taking into accounts these data one can suppose that the regions of the structural inhomogeneity that are outlined by the ellipses in Fig. 6a correspond to the walls of the tunnels of the matrix BC. The distance between them corresponds to the diameter of the tunnel. These peculiarities of the hydrogel's structure cannot be seen clearly, with high acutance in the SEM image because the BC ribbons which form these walls are surrounded by PAAm, the main component of the IPN. This component of the material screens the BC ribbons and scatters the electron beam during the SEM examination of the material.

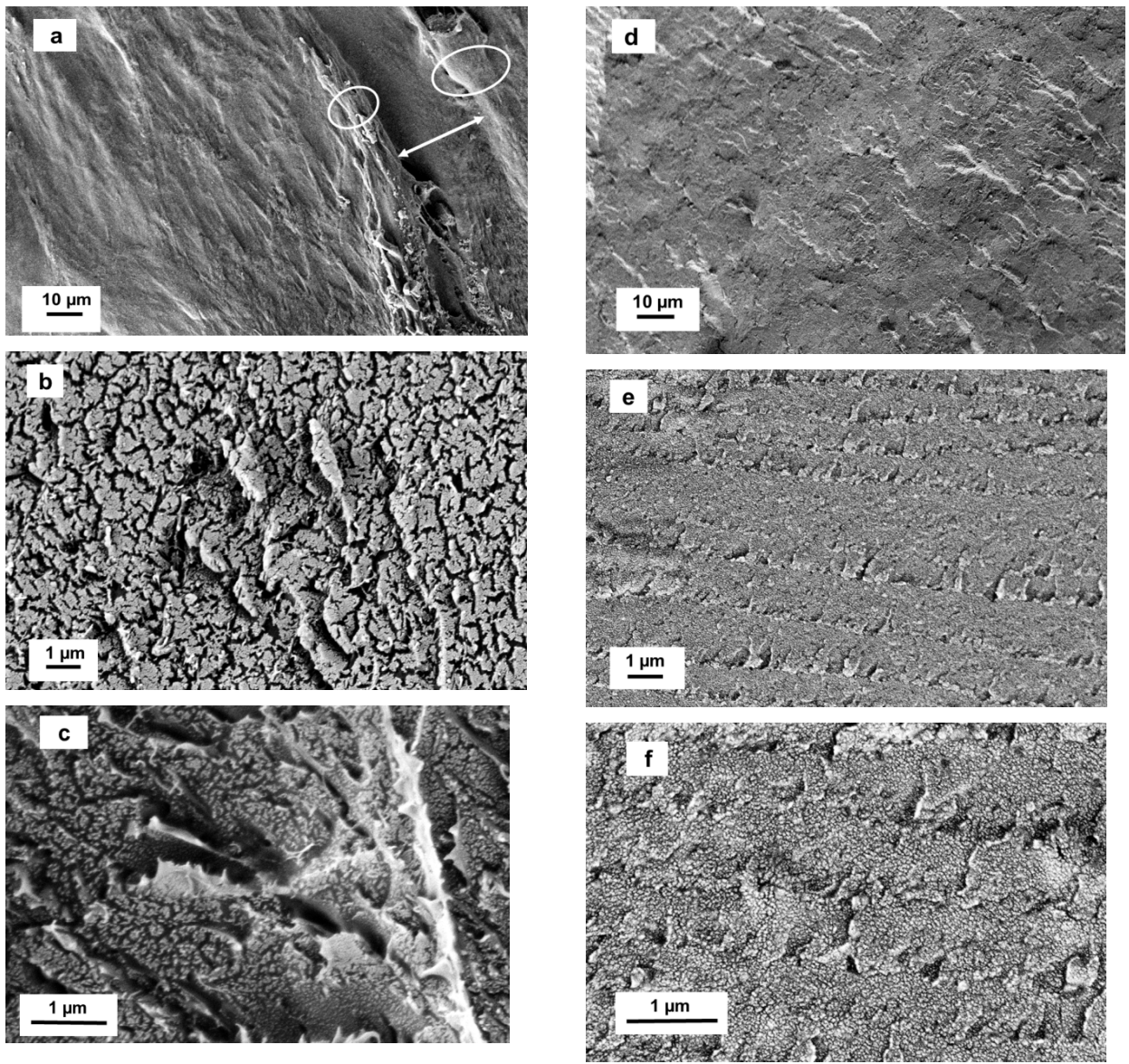

Fig. 6. Cryo-SEM micrographs of composite BC-PAAm (a, b, c) and PC-PAAm (d, e, f) hydrogels containing approx. 70 and 68 wt. % of water, respectively, with different magnifications: a – 2000x, b – 20000x, c – 40000x, d – 2000x; e – 20000x, f - 50000x.

At higher magnification (20 000x and 40 000x, Fig. 6 b and c, respectively) the domains with a size of about 1 μm can be clearly seen in the IPN structure of BC-PAAm hydrogels. Even smaller regions with dimensions of tens of nanometers are noticeable in the micron size domains. It is apparent that the borders between these domains arise due to the well-known effect of microphase separation between the polymeric IPN components (Sperling, 1994). For the system under investigation, it is possible to expect the emergence of an interface between macromolecules of PAAm and cellulose. So, the BC-PAAm hydrogels are characterized by a rather complicated hierarchy of the network architecture (or topology) that, undoubtedly, significantly affects the deformational behavior of these materials.

The structure of PC-PAAm hydrogels (Figs. 6 d – f) is substantially more homogeneous as compared to that of BC-PAAm gels. In PC-PAAm hydrogels no heterogeneities can be found of the characteristic dimensions of tens microns (Fig. 6 d), that agrees well with the specific peculiarities of the protocol of PC-PAAm hydrogels' synthesis.

The micro-photos of PC-PAAm gels with the magnifications 20 000x and 50 000x (Fig. 6 e and f, respectively), like those of BC-PAAm gels, demonstrate the presence of some domains of a micron size, but the interfaces, borders of these domains are less sharp. Really, the enhanced extent of homogeneity of structure could be expected to be realized in these gels based on PC, because in the synthesis of these materials the regenerated type of cellulose was used, that was obtained by the dissolution of cotton linter in trifluoroacetic acid. During this dissolution the fibrillar structure of cellulose fibers got broken. But it should be taken into account that in accordance with the data presented in (Roder, Morgenstern, 1999, Fink et al., 2001) after the cellulose dissolution some residual fragments of unsolved crystalline regions of substantial dimensions (up to 1000 macro-chains) can be detected in the solution. One can suppose that namely these crystalline fragments in the reactive solution during the synthesis of PC-PAAm gels originate the formation of the domains that are seen in Figs. 6 e and f.

Concerning the impact of the mesostructure, evidenced by Cryo-SEM, upon the mechanical properties of the hydrogels of both types it should be noted that, as it was stressed above, the anisotropy of the behavior inherent to the BC-PAAm hydrogels originates from the presence of channel-like structures oriented mainly in the direction transversal to the surface of the BC matrix layer; no such structures were detected while examining the structure of PC-PAAm gels.

The "walls" of these channel-like structures consist of the ribbons, the fibrillar bands composed of rigid BC macrochains, and namely these structures bear the mechanical load during the compression of the material in the direction parallel to the axis of above mentioned channels (perpendicular to the surface of BC cultivation). At the expense of a substantially high rigidity of cellulose macrochains these "walls" should possess a modest ability to withstand the elevated deformations without the breaking of the covalent bonds, at least – a partial rupture. Basing on the data above (Figs. 2, 4-5), these effects take place at the compressive deformations beyond 30-40 %. It was shown that namely at the compression levels of about 30 % the hydrogels of both types can bear the long-term cyclic loadings without any appearance of the progressive fall of mechanical stiffness. The increase of the compression amplitude up to 50 % causes the irreversible changes of the structure of the material along with a substantial decrease of mechanical stiffness (this effect was registered to take place in BC-based gel after several days of relaxation of the sample in water in unloaded state – Fig. 4a). Thus, under the action of the compression of 30 % the effects of the rearrangement of the IPN topology predominantly take place in these materials, while the increase of the compression deformation up to 50 % provokes the partial break of the covalent bonds that occurs most likely in the cellulose chains. The substantial intensification of these processes of the destruction of the system of covalent bonds takes place if the compression amplitude is increased up to 70 – 80% (Fig. 5). These phenomena lead to the dramatic decrease of the stiffness of the BC-based material in the wide range of deformation values. The further dramatic increase of the compressive stress registered in the compression range of 60-80 % can, presumably, be caused by the orientation of the micro-fibrillar structure of BC. Indeed, while treating these phenomena the fact should be taken into account that was evidenced in (Miquelard-Garnier et al., 2008): under the action of these high compressive deformations, the compression should take place of the molecular fragments of the network oriented parallel to the direction of the deformation, and, at the same time, the extension of the fragments oriented in the perpendicular direction. And it was shown in (Astley et al., 2001) that the effects of the orientation of the micro-fibrillar structure of the gel-pellicle of BC really take place in the course of its extension.

### Conclusion

The PAAm-based composite hydrogels, reinforced by both bacterial and plant cellulose are characterized by a high stiffness in their initial state ($E|_{10-15\%} \approx 8 - 9$ MPa in the first compression). Moreover, no appearance of visually seen destruction of both types of the materials was registered after the action of extremely severe compressions: up to 80 %. In the conditions of successive compressions the stress-strain behavior of these two types of hydrogels differs significantly. This effect is caused by the substantial differences in the structures and morphologic features of the reinforcing cellulose components of the interpenetrating networks of these two types.

In accordance with the SEM data both types of hydrogels studied are characterized by the microheterogeneity of structure caused by the realization of the microphase separation of both polymer components of IPN. Nevertheless, the extent of this effect of the phase separation in the BC-PAAm hydrogel substantially exceeds that in the PC-PAAm material. This difference is apparently provoked by the inhomogeneous microfibrillar organization of the physical network of BC.

The improved homogeneity of the distribution of stiff cellulose macrochains in the hydrogels reinforced by the regenerated plant cellulose insures the improved stability of their behavior under the action of high multiple and cyclic compressive loadings. Indeed, the amplitude values of compressive strength for these hydrogels are as high as 2 – 8 MPa (under the action of compressive deformations with the amplitude from 30 to 70 %) and no marked decrease of the stress values was registered in the long term cyclic tests yet in the conditions of the high compression amplitudes (up to 70 %). This behavior differs significantly from that of BC-based hydrogels: they cannot withstand in their initial structural state the action of cyclic compression with the amplitude more than 30 %; the appropriate value of maximal stress does not exceed 2 MPa.

The complex hierarchy of supramolecular structure of BC causes some specific effects that take place in the processes of deformation of hydrogel compositions based on this type of cellulose. The oriented tunnel-like structures with the characteristic dimensions of 10 μm and more that exist in this type of cellulose cause the anisotropic character of the mechanical characteristics of BC-PAAm hydrogels.

A great increase of the compressive stress registered while compressing the BC-PAAm hydrogels in the deformation range beyond 60 % (for the samples that were previously compressed up to 80 %) can presumably be provoked by the reorientation of cellulose microfibrils under the action of this extensive compression; in these conditions they can be stretched in the direction along the axes perpendicular to the compression direction.

The PC-based hydrogels are optimal to be used as the artificial substituent of cartilage. The hydrogels reinforced by BC are less suitable to gain this purpose. Nevertheless, they apparently could be used to the repair the zones of articular cartilage that are not subjected to the action of high compression stresses.

Of the particular interest is the anisotropy of the mechanical properties inherent to BC-PAAm hydrogels. This specific feature is similar to that registered while studying the compressive behavior of articular cartilages (Jurvelin et al., 2003). Under this reason it would be interesting to try to modify the stricture of these hydrogels by different methods with the goal of further improvement of their mechanical properties.

The BC-PAAm hydrogels can also be treated as the promising precursors of bone tissues because, being subjected to a contact with bones in the organism, these materials are involved in the extensive mineralization process that yields the formation of calcium phosphate spherulites in the hydrogel volume. The chemical composition of this salt is close to that of hydroxyapatite (Buyanov et al., 2016). This process can result in the formation of a composite material close to bones by its mechanical properties.

### Acknowledgements

The authors are grateful to Dr. A.A. Tkachenko and Dr. A.K. Khripunov for providing the samples of bacterial cellulose.

The work of Dr. I.V. Gofman was supported financially by the Ministry of Education and Science of the Russian Federation (state contract No. 14.W03.31.0014).